\documentclass[titlepage,revtex,12pt]{article}
\usepackage{amsmath,amssymb}
\begin{document}

\begin{titlepage}
\thispagestyle{empty}
\begin{center}
{\Large\bf Moving system with speeded-up evolution}\\[0.5cm]
{\large\bf M. I. Shirokov }\\[0.5cm]
{Bogoliubov Laboratory of Theoretical Physics\\
Joint Institute for Nuclear Research\\
141980 Dubna, Russia\\
e-mail: shirokov@theor.jinr.ru}
\end{center}

\vspace*{1cm} \noindent
{\large Abstract}\\

In the classical (non-quantum) relativity theory the course of the moving clock
is dilated as compared to the course of the clock at rest (the Einstein dilation).
Any unstable system may be regarded as a clock. The time evolution (e.g., the decay)
of a uniformly moving physical system is considered using the relativistic
quantum theory. The example of a moving system is given whose evolution turns out
to be speeded-up instead of being dilated. A discussion of this paradoxical
result is presented.
\end{titlepage}


\section{INTRODUCTION}
\label{intro}

The classical (nonquantum) relativistic theory states that the course of the
moving clock is dilated as compared to the course of the clock at rest (the
Einstein dilation (ED)), e.g., see \cite{1}. Any nonstationary physical system
(e.g., an excited hydrogen atom) may be regarded as a clock \cite{1}. Here the
time evolution of a moving physical system is considered using the relativistic
quantum theory.

The decay of the moving unstable particle was examined in the papers
\cite{2,3,4,5}. The moving system was described by a state with a sharp nonzero
momentum. Its evolution was found to be consistent with ED up to high
precision. In \cite{6} I pointed out an example of the nonstationary state
whose evolution is speeded-up instead of being dilated. This curious result is
the exact consequence of the relativistic quantum theory.
By definition this theory must contain operators $H$ and $\vec{P}$ of total
energy and momentum ( the Lee group generators of time and space translations),
total angular momenta, and generators of the Lorentz boosts $\vec{N}$.
These generators must satisfy commutation relations of the Poincar\'e group.
Usual Dirac's ``instant form" of the theory is implied in which time evolution
is described by the operator $\exp (-iHt)$. Note that in this form interaction
terms are contained in $\vec{N}$ along with $H$
\begin{equation}
\label{eq1}
H=H_0+H_{\rm int}, \qquad \vec{N}=\vec{N}_0+\vec{N}_{\rm int}.
\end{equation}
Indeed, the commutation relation $[N_i,P_j]=i\delta_{ij}H$ of the
Poincar\'e group means that if $P$ does not contain interaction terms,
then $\vec{N}$ must contain them along with $H$.

A simple example of a theory like that may be the Lee model of the decay of an unstable
particle $a$ into stable particles $b$ and $c$: $a\rightarrow b+c$. The interaction
terms are of the threelinear kind : $\hat{a}\hat{b}^\dag\hat{c}^\dag +{\rm h.c.}$
(momentum indices of destruction-creation operators are omitted).

The moving unstable particle state is described in \cite{3}-\cite{5} by the
vector $a_p^\dag\Omega_0$, $a_p^\dag$ being the creation operator of the particle
$a$ with momentum $p$. When $p=0$ the vector $a_0^\dag\Omega_0$ describes the state
``one unstable particle at rest, no decay products".

In \cite{6}, the moving particle was described by the vector $\Phi_v =L_v a_0^\dag\Omega_0$,
$L_v$ being the Lorentz transformation (see Eq.~(\ref{eq6}) below) from the frame where
the particle velocity is zero to the frame where the particle velocity is equal to $v$.
The state $\Phi_v$ differs from $a_p^\dag\Omega_0$. Indeed, $L_v$ contains interaction.
Therefore, $\Phi_v =L_v a_0^\dag\Omega_0$ has an admixture of decay particles
(if $v\neq 0$) in analogy with the state $\exp (-iHt)a_0^\dag\Omega_0$. So $\Phi_v$
is not a pure ``one-unstable-particle-state", it contains decay products.

The system state $\Phi_v$ along with $a_p^\dag\Omega$ is nonstationary and the system
may be considered as a quantum clock.

The evolution of the state $\Phi_v$ was compared in \cite{6} with the evolution
of the system at rest. It turns out that the former is speeded-up as compared to the
latter, not dilated. G.~Hegerfeldt in \cite{7} pointed out a simple way of the
derivation of this curious fact. I suppose that the detailed presentation of this way
is justified. It is set forth in Sect.~\ref{s2} and Appendix. I use a modified initial
state $\Phi_v$ as compared to \cite{6} and \cite{7}, and characterize its evolution
by a different amplitude. These modifications allow us to meet the Hegerfeldt critical
remarks, see \cite{7}, p.~208. The result is discussed in Conclusion.


\section{Moving system with speeded-up evolution}
\label{s2}

Consider the scalar product
\begin{equation}
\label{eq2}
V(v,t)=\langle L_v\varphi_0,\exp (-iHt)L_v\Phi_0\rangle .
\end{equation}
Here $\varphi_0$ is a non-normalizable eigenvector of the total system momentum
having zero eigenvalue: $\vec{P}\varphi_0=0$. The vector $\Phi_0$ describes
a normalized packet having zero average momentum
$$
\langle\Phi_0,\Phi_0\rangle =1, \qquad \langle\Phi_0,\vec{P}\Phi_0\rangle =0.
$$

In \cite{6} and \cite{7} the akin scalar product
\begin{equation}
\label{eq3}
\langle L_v \varphi_0,\exp (-iHt)L_v\varphi_0\rangle
\end{equation}
was considered. Being a survival amplitude it has the following deficiency: it is a
scalar product of two non-normalized vectors and has no physical meaning. In particular,
at $t=0$ Eq.~(\ref{eq2}) turns into $\langle\varphi_0,\varphi_0\rangle =\infty$,
which is inadmissible for the amplitude of probability (the latter cannot exceed the unit).
Meanwhile $V(v,t)$ is the scalar product of non-normalizable and normalizable vectors.
It is not a survival amplitude: $L_v\varphi_0$ differs from the initial state $L_v\Phi_0$.
However, $V(v,t)$ may be endowed the meaning of a probability amplitude of finding
(detecting) the state $L_v\varphi_0$ in the state $\exp (-iHt)L_v\Phi_0$, see \cite{8},
v.~I, ch.~V, sect.~10.

Note that the amplitude~(\ref{eq3}) makes sense if the momentum would have a discrete
spectrum so that $\langle\varphi_0,\varphi_0\rangle =1$. This would be the case if
the system is implied to be in a large but finite space volume and usual periodicity
conditions are imposed (or the volume opposite boundaries are identified).
However, a discrete momentum spectrum is not consistent with the Lorentz transformation
law which will be essentially used below, see Eqs.~(\ref{eq5}) and (\ref{eq11}).

Using the formula $L^{-1}f(H)L=f(L^{-1}HL)$ (where $f$ is an exponential) one can
represent (\ref{eq2}) as
\begin{equation}
\label{eq4}
V(v,t)=\langle\varphi_0,L_v^\dag\exp (-iHt)L_v\Phi_0\rangle
=\langle\varphi_0,\exp (-itL_v^\dag HL_v)\Phi_0\rangle .
\end{equation}
Here $L_v^\dag HL_v$ is the Lorentzian-transformed Hamiltonian. If $H$ and $\vec{P}$
were $c$-numbers, then the transformed energy would have the known expression
in terms of initial energy and momentum:
$$
H'=H\gamma -\vec{P}\vec{v}\gamma, \qquad \gamma =(1-v^2)^{-1/2}.
$$
Intuitively one may expect that a similar equation holds for operators $\hat{H}$, $\hat{\vec{P}}$:
\begin{equation}
\label{eq5}
L_v^\dag \hat{H}L_v =\hat{H}\gamma -\hat{\vec{P}}\vec{v}\gamma .
\end{equation}
This conjuncture may be confirmed by an algebraic calculation with
\begin{equation}
\label{eq6}
L_v =\exp i\vec{\beta}\vec{N}, \qquad \tanh |\vec{\beta}|=|\vec{v}|,
\qquad \vec{\beta}\|\vec{v}
\end{equation}
(see \cite{9}). This calculation is carried out in Appendix.

In what follows let us assume $\vec{v}=(0,0,v)$, $\vec{\beta}=(0,0,\beta)$
and that $N$, $\hat{P}$ denote, respectively, $N_3$, $\hat{P}_3$.
Now we may continue Eq.~(\ref{eq4}):
\begin{eqnarray}
\label{eq7}
V(v,t)&=&\langle\varphi_0,\exp \left[ -it(\hat{H}\gamma -\hat{P}v\gamma)\right]
\Phi_0\rangle \nonumber\\
&=&\langle\varphi_0,\exp (it\hat{P}v\gamma)\exp (-it\hat{H}\gamma)\Phi_0\rangle\\
&=&\langle\varphi_0,\exp (-it\hat{H}\gamma)\Phi_0\rangle .\nonumber
\end{eqnarray}
Here the equation $\exp (A+B)=\exp B \exp A$ has been used which is valid for commuting
operators $A=-it\hat{H}\gamma$ and $B=it\hat{P}v\gamma$. As $\varphi_0$ is the $\hat{P}$
eigenvector corresponding to zero eigenvalue, we have
$\exp (-it\hat{P}v\gamma)\varphi_0=\varphi_0$. So we obtain
\begin{equation}
\label{eq8}
V(v,t)\equiv\langle L_v\varphi_0,\exp (-itH)\Phi_v\rangle
=\langle\varphi_0,\exp (-itH\gamma)\Phi_0\rangle .
\end{equation}
Compare the result with the amplitude $V(v,t)$ at $v=0$ (when $L_v=1$). One obtains
\begin{equation}
\label{eq9}
V(v,t)=V(0,t\gamma).
\end{equation}
This equation is an \underline{exact} expression of $V(v,t)$ in terms of $V(0,t)$.
I do not intend and need to calculate $V(0,t)$ (for this calculation see \cite{2}).
As $\gamma =(1-v^2)^{-1/2}\geq 1$, relation (\ref{eq9}) shows that a moving system evolves
faster than the system at rest: the amplitude $V(v,t)$ at the moment $t$ assumes the
value that $V(0,t)$ assumes at a later moment $\gamma t$.

For the corresponding probabilities one obtains from Eq.~(\ref{eq9}) the equation
\begin{equation}
\label{eq10}
|V(v,t)|^2=|V(0,t\gamma)|^2 .
\end{equation}

It is of interest to consider such properties of the state $L_v\varphi_0$
as its momentum and energy.

One may show that $L_v\varphi_0$ is not an eigenvector of the total momentum $\hat{P}$.
For this purpose the equation
\begin{equation}
\label{eq11}
{\rm e}^{-i\beta N}\hat{P}{\rm e}^{i\beta N}=\gamma\hat{P}-\gamma v\hat{H},
\qquad \tanh\beta =v
\end{equation}
may be used. It is the Lorentz transformation of the momentum operator, cf.~(\ref{eq5}).

Using Eq.~(\ref{eq11}) one may calculate average momentum $\langle P\rangle_v$
of the state $L_v\varphi_0$. One obtains
\begin{equation}
\label{eq12}
\langle P\rangle_v\equiv\langle L_v\varphi_0,\hat{P}L_v\varphi_0\rangle
=\gamma v\langle\varphi_0,\hat{H}\varphi_0\rangle .
\end{equation}
Here $\langle\varphi_0,\hat{H}\varphi_0\rangle$ is the average energy of the state
$\varphi_0$. Note that $\varphi_0$ is not $\hat{H}$ eigenvector: $\varphi_0$
describes an unstable particle and is a nonstationary state.

Using Eq.~(\ref{eq5}) one may calculate the average energy $\langle E\rangle_v$
of the state $L_v\varphi_0$
\begin{equation}
\label{eq13}
\langle E\rangle_v\equiv\langle L_v\varphi_0,\hat{H}L_v\varphi_0\rangle
=\gamma\langle\varphi_0,\hat{H}\varphi_0\rangle .
\end{equation}
It follows from Eqs.~(\ref{eq12}) and (\ref{eq13}) that $\langle P\rangle_v$ and
$\langle E\rangle_v$ satisfy the relativistic relation
$\langle P\rangle_v =v\langle E\rangle_v$.


\section{Time evolution of a moving unstable particle with exact momentum}
\label{s3}

In \cite{3}, \cite{4}, \cite{5}, the state of a moving unstable particle was
described by the eigenvector $\psi_p$ of the momentum $\hat{P}$:
$\hat{P}\psi_p=p\psi_p$ (if $\psi_p$ describes one unstable particle, then the total
momentum coincides with particle momentum). I assume that the momentum spectrum is
discrete (see sect.~\ref{s2}) and consider the survival amplitudes
\begin{eqnarray}
\label{eq14}
A_p(t)=\langle\psi_p,\exp (-itH)\psi_p\rangle ,\\
\label{eq15}
A_0(t)=\langle\psi_0,\exp (-itH)\psi_0\rangle .
\end{eqnarray}
Now the amplitude $A_p(t)$ is not connected with $A_0(t)$ by such a simple relation
as $V(v,t)$ and $V(0,t)$ do. To compare $A_p(t)$ with $A_0(t)$, one has to calculate
them separately. Let us write out from \cite{5} approximates expressions for
$A_p(t)$ and $A_0(t)$ which are valid for time not too short and not too long
(when the decay laws are exponential)
\begin{eqnarray}
\label{eq16}
A_0(t)&\cong&\exp (-imt-\Gamma t/2), \\
\label{eq17}
A_p(t)&\cong&\exp (-imt\gamma_m-\Gamma t/2\gamma_m),
\qquad \gamma_m =\sqrt{p^2+m^2}/m.
\end{eqnarray}
Here $m$ is the average (or most probable) mass $\langle\psi_0,H\psi_0\rangle$
of the unstable particle. It follows from Eqs.~(\ref{eq16}) and (\ref{eq17})
that
\begin{equation}
\label{eq18}
|A_p(t)|^2\cong |A_0(t/\gamma_m)|^2 .
\end{equation}
Note that $\gamma_m$ coincides with $(1-v^2)^{-1/2}$ at $v=P/\sqrt{P^2+m^2}$.
Equation~(\ref{eq18}) means that the Einstein dilation holds. This must be juxtaposed
to the speeding-up expressed by Eq.~(\ref{eq10}).


\section{CONCLUSION}
\label{con}

When deriving the Einstein dilation (ED) in the classical (nonquantum) relativity theory,
one uses the notion of a clock which may simultaneously have the exact (e.g., zero) velocity
and be in a definite position (e.g., in the coordinate origin). This is impossible for
a quantum clock.

Any nonstationary (unstable) quantum system may be regarded as a quantum clock. The time
evolution of the moving system is here considered by using the relativistic quantum
theory, see Introduction. The Lorentzian transformation of coordinates and time is not used.
Instead, the transformation of momenta and energy is exploited, see Eqs.~(\ref{eq11})
and (\ref{eq5}).

It was earlier shown in \cite{3}-\cite{5} that ED holds approximately when the moving quantum
system has definite momentum, see Eq.~(\ref{eq18}). Here I present another description of
the moving system. Its time evolution turns out to be speeded-up instead of being dilated.

So the system evolution depends upon a choice of the initial state of the system.

The obtained relation, Eq.~(\ref{eq9}), between the laws of evolution of the moving system
and the system at rest is the exact corollary of the used quantum postulates and of the assumed
description of the moving system. Equation~(\ref{eq9}) does not depend on a concrete interaction
guiding the evolution.

Quantum postulates do not forbid the existence of the state $\Phi_v$ which gives speeding-up.
However, experiments agree with ED. The used quantum theory may explain this fact assuming
that the state of the measured unstable system is of the kind $\psi_p$, see sect.~\ref{s3}.

It is more natural to suppose that the initial state prepared in a real experiment is not
the state ``unstable particle, no decay products" but rather the state ``unstable particle
together with decay products (the background)". The latter may be assumed to be a
superposition of states $\psi_p$ and $\Phi_v$. When the $\gamma$-factor is sufficiently large,
the short-lived component $\Phi_v$ dies out as time grows. Only the long-lived component
$\psi_p$ survives. In other words, experiments tend to detect the long-lived component.
So the measured life-time of such a moving system may depend on a measuring device.


\vspace*{0.4cm}
\begin{center}
{\bf ACKNOWLEDGMENTS}
\end{center}
I am grateful to Prof. G. S. Hegerfeldt for his comment on my paper \cite{6}.



\newpage
\renewcommand{\theequation}{A.\arabic{equation}}
\setcounter{equation}{0}  
\begin{center}
{\bf APPENDIX}
\end{center}
\vspace*{0.1cm}

The derivation of the main result (\ref{eq10}) is based on the equation
\begin{equation}
\label{a1}
\exp (i\vec{\beta}\vec{N})H\exp (-i\vec{\beta}\vec{N})=H\gamma -\vec{P}\vec{v}\gamma .
\end{equation}
The equation will be proved by using the method of ``parameter differentiation", e.g.,
see \cite{10}, sect.~6, p.~969.

Suppose that the vector $\vec{v}=(v_1,v_2,v_3)$ is directed along the $z$-axis so that
$v_1=v_2=0$. Note that $\vec{\beta}$ is parallel to $\vec{v}$. Then
$\exp (i\vec{\beta}\vec{N})=\exp (i\beta_3N_3)$. In what follows $v$, $\beta$, $N$, $P$
denote, respectively, $v_3$, $\beta_3$, $N_3$, $P_3$. Three operators $N$, $H$, $P$ form
the subalgebra of the Poincar\'e generators:
\begin{equation}
\label{a2}
[N,H]=iP, \qquad [N,P]=iH, \qquad [H,P]=0.
\end{equation}
Any element of this subalgebra may be expanded over $N$, $H$, $P$. So does, in particular,
the element $H_\beta =\exp (i\beta N)H\exp (-i\beta N)$:
\begin{equation}
\label{a3}
H_\beta =h(\beta)H+p(\beta)P+n(\beta)N.
\end{equation}
Here $h(\beta)$, $p(\beta)$, and $n(\beta)$ are $c$-number functions of $\beta$. Using the
well-known expansion
$$
{\rm e}^A B{\rm e}^{-A}=B+[A,B]+\frac{1}{2!}[A,[A,B]]+\ldots
$$
one may verify that $n(\beta)$ in Eq.~(\ref{a3}) is zero. We have
\begin{equation}
\label{a4}
dH_\beta /d\beta =Hdh(\beta)/d\beta +Pdp(\beta)/d\beta .
\end{equation}
On the other hand,
\begin{eqnarray}
\label{a5}
dH_\beta/d\beta &=&d\left( {\rm e}^{i\beta N}H{\rm e}^{-i\beta N}\right)/d\beta
=iN{\rm e}^{i\beta N}H{\rm e}^{-i\beta N}-i{\rm e}^{i\beta N}H{\rm e}^{-i\beta N}N
\nonumber\\
&=& i[N,H_\beta]=i[N,hH+pP]=-hP-pH
\end{eqnarray}
(the commutator relations~(\ref{a2}) were used). Equating the r.h.s. of (\ref{a4})
and (\ref{a5}) we obtain
\begin{equation}
\label{a6}
Hdh(\beta)/d\beta +Pdp(\beta)/d\beta =-Ph(\beta)-Hp(\beta).
\end{equation}
Since $H$ and $P$ are independent operators, we must have
\begin{equation}
\label{a7}
dh(\beta)/d\beta +p=0, \qquad dp(\beta)/d\beta +h(\beta)=0.
\end{equation}
This is the system of ordinary differential equations. Its solution may be found in
the book~\cite{11}, ch.~VIII, sect.~8.3:
\begin{equation}
\label{a8}
h(\beta)=\cosh\beta, \qquad p(\beta)=-\sinh\beta.
\end{equation}
So we get from Eqs.~(\ref{a3}) and (\ref{a8})
\begin{equation}
\label{a9}
{\rm e}^{i\beta N}H{\rm e}^{-i\beta N}=H\cosh\beta -P\sinh\beta .
\end{equation}
As $\tanh\beta =v$, we have
\begin{equation}
\label{a10}
\cosh\beta =(1-\tanh^2\beta)^{-1/2}=(1-v^2)^{-1/2}=\gamma, \qquad \sinh\beta =v\gamma.
\end{equation}
So Eq.~(\ref{a9}) may be rewritten as Eq.~(\ref{eq5}).




\end{document}